\documentclass[12pt]{aastex} 
\usepackage[twocolumn]{emulateapj5}
\usepackage{epic}

\newcommand{\hrefkpc} {$h_{70}^{-1}$\,kpc}
\newcommand{\lcdm} {$\Lambda$CDM }
\newcommand{\hrefmpc} {$h_{70}^3$ Mpc$^{-3}$}
\gdef\kms{km\,s$^{-1}$}
\gdef\1054{MS\,1054$-$03}
\gdef\2053{MS\,2053$-$04}
\gdef\omit#1{}

\shortauthors{Labb\'e et al.}
\shorttitle{Large disk-like galaxies at high redshift}
\slugcomment{}

\def\figa{
\begin{figure*}[t]
\includegraphics[width=19.cm,bb=32 186 620 606]{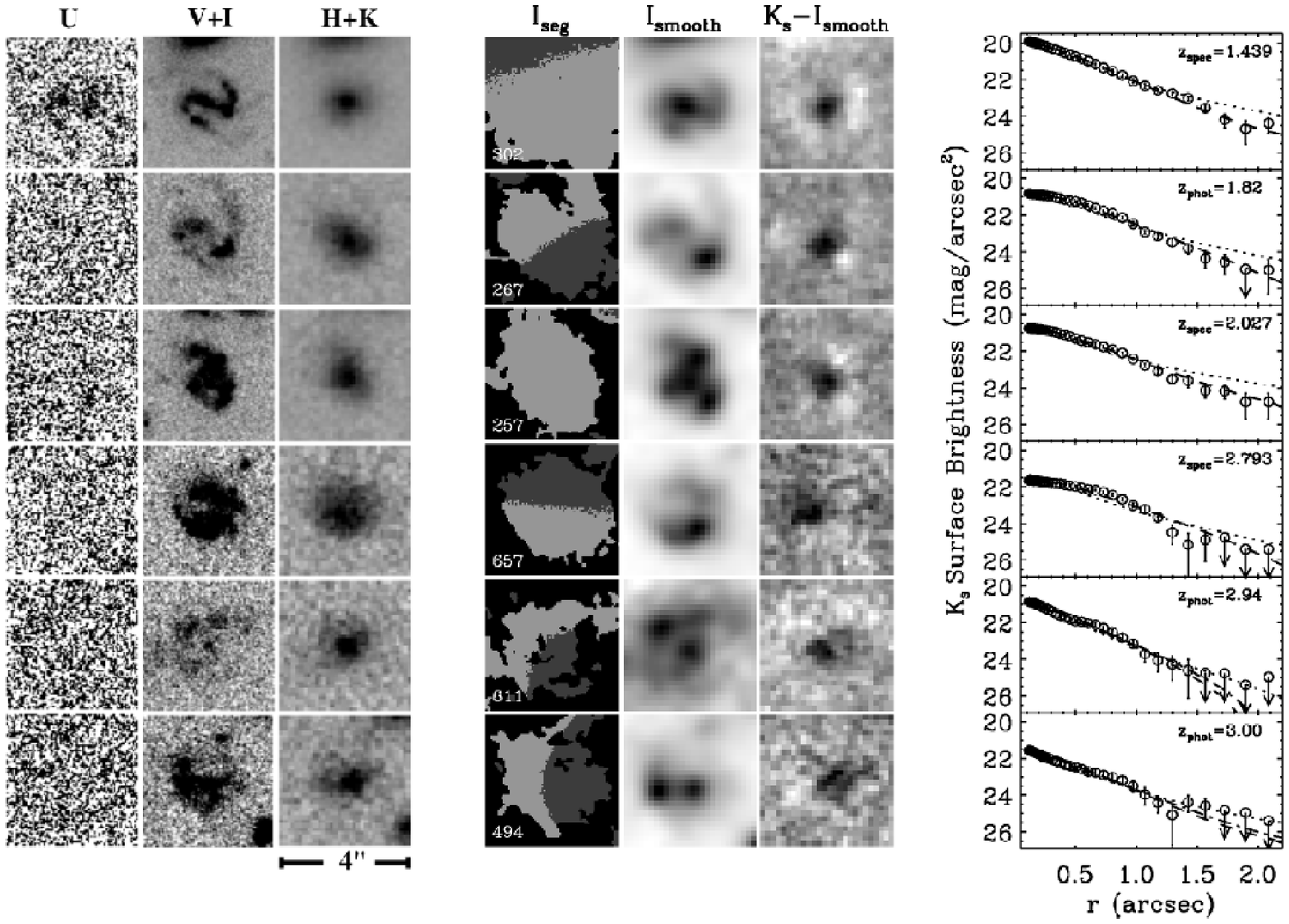}
\figcaption{\small 
{\it Left panels:} The WFPC2 $U_{300}$, averaged $V_{606}+I_{814}$ (rest-frame UV), 
and our averaged ISAAC $H+K_s$ images (rest-frame optical), scaled proportional to 
$F_\lambda$ with arbitrary normalization per galaxy. 
The rest-frame UV morphologies are complex and symmetric with respect to the center
of the smoother optical distribution.  {\it Middle panels:} SExtractor's  
$I_{814}$-band segmentation map, the smoothed $I_{814}$ images, and the $K_s$-band images after subtracting the scaled, smoothed $I_{814}$-band. From the segmentation map it follows
that detection in $I_{814}$ would likely split up most of the sources. The central residuals in $K_s-I_{smooth}$ demonstrate that optical and NIR light are distributed differently
and that all galaxies have a ``red'' nucleus.
{\it Right panels:} The radial profiles in the $K_s$-band. The abscissa and the 
ordinate are respectively the mean geometric radial distance and the surface 
brightness along elliptical isophotes. The arrows mark $1\sigma$ confidence intervals 
for measurements with signal-to-noise less than 1. Overplotted are the best-fit 
exponential law ({\it dashed}), $r^{1/4}$ law ({\it dotted}), and point+exponential 
({\it dash-dot}) for galaxy 494 and 611. 
\label{fig.a}}
\end{figure*}

}

\def\figb{
\leavevmode
\vspace{-0.5cm} 
\begin{center}
  \vbox{
      \includegraphics[width=9.cm,height=3.3cm, bb=128 113 468 240]{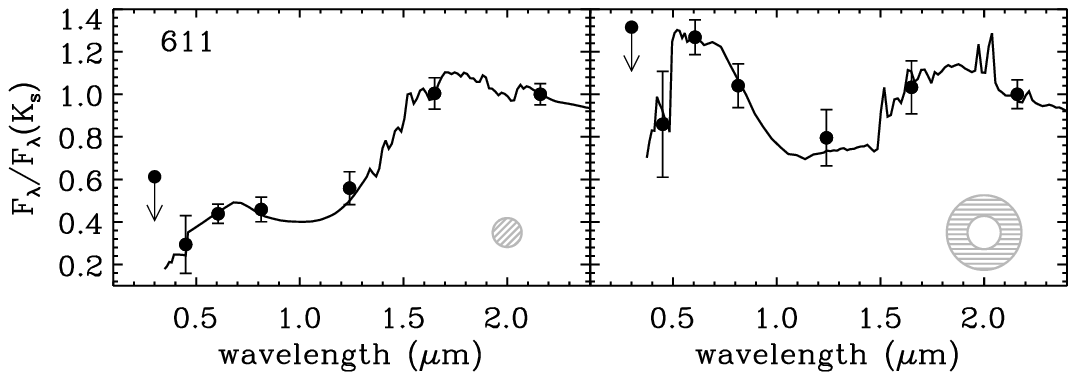}   
  }

\figcaption{\small
  The spectral energy distribution of source 611
  in a $0\farcs7$ circular diameter aperture ({\it left}) and in a
  concentric $1\arcsec - 2\arcsec$ diameter ring ({\it right}), normalized to 
  the $K_s$-band flux. Overplotted are independent model fits from 
\citet{Ru03a}.
\label{fig.c}}
  \end{center}
}

\begin{document}
\title{Large disk-like galaxies at high redshift\altaffilmark{1}}
\author{
Ivo Labb\'{e}\altaffilmark{2}, 
Gregory Rudnick\altaffilmark{3}, 
Marijn Franx\altaffilmark{2},
Emanuele Daddi\altaffilmark{5},
Pieter G. van Dokkum\altaffilmark{6}, 
Natascha M. F\"{o}rster Schreiber\altaffilmark{2}, 
Konrad Kuijken\altaffilmark{2}, 
Alan Moorwood\altaffilmark{5}, 
Hans-Walter Rix\altaffilmark{4}, 
Huub R\"{o}ttgering\altaffilmark{2}, 
Ignacio Trujillo\altaffilmark{4},
Arjen van de Wel\altaffilmark{2}, 
Paul van der Werf\altaffilmark{2}, 
Lottie van Starkenburg\altaffilmark{2}
}

\begin{center}
{\em Accepted for publication in the Astronomical Journal Letters}
\end{center}

\altaffiltext{1}{Based on service mode observations collected at 
the European Southern Observatory, Paranal, Chile 
(ESO Programme 164.O-0612). Also based on observations with the 
NASA/ESA {\em Hubble Space Telescope}, obtained at the Space 
Telescope Science Institute which is operated by AURA, Inc., 
under NASA contract NAS5-26555.
}

\altaffiltext{2}{Leiden Observatory, P.O. Box 9513, NL-2300 RA,
Leiden, The Netherlands}

\altaffiltext{3}{Max-Plank-Institut f\"ur Astrophysik, Postfach 1317,
D-85741, Garching, Germany}

\altaffiltext{4}{Max-Plank-Institut f\"ur Astronomie, D-69117,
Heidelberg, Germany }

\altaffiltext{5}{European Southern Observatory, D-85748, Garching,
Germany }

\altaffiltext{6}{Department of Astronomy, Yale University, P.O. Box 208101, 
New Haven, CT 06520-8101}

\begin{abstract}
Using deep near-infrared imaging of the Hubble Deep
Field South with ISAAC on the {\em Very Large Telescope} 
we find 6 large  disk-like galaxies at redshifts $z = 1.4-3.0$. 
The galaxies, selected in $K_s$ ($2.2\mu m$), are regular and surprisingly 
large in the near-infrared (rest-frame optical), with face-on effective radii
$r_e=0\farcs65-0\farcs9$ or $5.0-7.5$ \hrefkpc\ in a \lcdm cosmology,
comparable to the Milky Way. The surface brightness profiles are 
consistent with an exponential law over $2-3$ effective radii. 
The WFPC2 morphologies in {\em Hubble Space Telescope} imaging (rest-frame 
UV) are irregular and show complex aggregates of star-forming 
regions $\sim 2\arcsec$ ($\sim15$ \hrefkpc) across, symmetrically
distributed around the $K_s$-band centers. 
The spectral energy distributions show clear breaks in the rest-frame
optical. The breaks are strongest in the central regions of the galaxies, 
and can be identified as the age-sensitive Balmer/4000\,\AA\ break.
The most straightforward interpretation is that these galaxies
are large disk galaxies; deep NIR data are indispensable for this classification.
The candidate disks constitute 50\% of galaxies with
$L_V \gtrsim 6\times 10^{10} h_{70}^{-2} L_{\sun}$ at $z = 1.4-3.0$.
This discovery was not expected on the basis of previously studied samples. In 
particular, the Hubble Deep Field North is deficient in large galaxies 
with the morphologies and profiles we report here.

\leavevmode
\vspace{0.cm} 

\end{abstract}

\section{Introduction}

\figa

Disk galaxies are believed to undergo a relatively simple formation process
in which gas cools and contracts in dark
matter halos to form rotationally supported disks with exponential light 
profiles \citep{FE80,Mo98}. 
A critical test of any theory of galaxy formation is to reproduce the
observed properties and evolution of galaxy disks.
\par
Previous optical spectroscopy and HST imaging have yielded a wealth 
of data on disk galaxies at $z\lesssim1.$ (e.g., \citealt{Vo96,Vo97,Li98,Ba03}), 
although contradictory claims have been made regarding the implications 
for the size and luminosity evolution with redshift 
(see \citealt{Li98, MMW98, Ba03}), and the importance of surface 
brightness selection effects (see \citealt{Si99,BS02}).
\par
It is still unknown what the space density and properties are of disk galaxies 
at substantially higher redshift. Many galaxies at $z\sim3$ have been identified 
using the efficient U-dropout technique \citep{S96a,S96b}. Most of these
objects are compact with radii $\sim 1-2$ \hrefkpc, while some are large and 
irregular \citep{GSM96,Lo97}. However, the U-drop selection requires high 
far-UV surface brightness due to active, spatially compact, and unobscured 
star formation. 
As a result, large and UV-faint disk galaxies may have
been overlooked and, additionally, the morphologies of LBGs could just reveal 
the unobscured star-forming regions rather than the more evolved 
underlying population which forms the disk. 
\par
The most direct evidence for the existence of large disks at high redshift 
has come from observations in the NIR, which provide access to the rest-frame
optical. Here the continuum light is more indicative of the distribution 
of stellar mass than in the UV and nebular lines are accessible for 
kinematic measurements. \citet{vDS01} discuss a 
$K$-selected galaxy at $z = 1.34$ with a rotation velocity of 
$\sim 290$~km~s$^{-1}$. \citet{Erb03} detect $\sim 150$~km~s$^{-1}$ 
rotation at $\sim 6$ \hrefkpc\ radii in the $H\alpha$ emission line of 
galaxies at $z \sim 2.3$ and \citet{Mo03} find $\gtrsim 100$~\kms\ rotation 
at $\sim 6$ \hrefkpc\ from the center of a galaxy at $z = 3.2$, seen in the NIR 
spectrum of the [\ion{O}{3}]$\lambda$5007 \AA\ emission line.
\par
The imaging data in these studies, however, are of limited depth and resolution, making it 
difficult to determine morphological properties. In this Letter, we present an analysis of the rest-frame ultraviolet-to-optical morphologies and spectral energy distributions 
(SEDs) of 6 large candidate disk galaxies at $z\sim1.4-3$ using the deepest 
groundbased NIR dataset currently available \citep{La03}.  Throughout, we adopt 
a flat $\Lambda$-dominated cosmology ($\Omega_M=0.3, \Lambda=0.7, H_0 = 70~h_{70}$~kms$^{-1}$Mpc$^{-1}$). All magnitudes are expressed in the Johnson 
photometric system. 

\begin{small}
\begin{table*}[t]
\begin{center}
\caption{Properties of high redshift disk galaxies in the HDF-S}
\begin{tabular}{lcllclccc}
\hline
\hline
Galaxy$^a$ & $K_{s,tot}^b$ & $z$ & $M_{B,rest}^c$ & $\mu_{0,B,rest}^d$& $r_{e,K}^e$ & $r_{1/2,K}^f$ & $r_{1/2,I}^g$ & $\epsilon^h$ \\ 
\hline
302 & 19.70  & 1.439$^i$  & -22.70 & 19.70 & 0.89  & 0.70 & 0.86 & 0.46\\
267 & 19.98  & 1.82  & -22.88 & 19.92 &  0.75  & 0.74 & 0.88 & 0.37\\
257 & 20.25  & 2.027$^i$ & -23.08 & 19.53 & 0.74  & 0.74 & 0.84 & 0.36\\
657 & 20.68  & 2.793$^i$  & -23.56 & 19.33 & 0.76  & 0.70 & 0.74 & 0.18 \\
611 & 20.53  & 2.94  & -23.59 & 18.51 & 0.65$^j$  & 0.52 & 0.97 & 0.27 \\
494 & 21.14  & 3.00   &  -23.31 & 18.84 & 0.75$^j$  & 0.56 & 0.86 & 0.47\\
\hline
\hline
\label{tab.a}
\end{tabular}
 \begin{small}
\begin{tabular}{ll}
$^a$\,Catalog identification numbers (see \citealt{La03})   & $^f$\,$K_s$ half-light radii (arcsec)\\
$^b$\,$K_s$-band total magnitudes                       & $^g$\,$I_{814}$ half-ligt radii, PSF-matched to $K_s$ (arcsec)\\
$^c$\,Rest-frame absolute $B$-band magnitudes     & $^h$\,Ellipticity \\
$^d$\,Face-on rest-frame $B$-band central surface brightnesses   &   $^i$\,Spectroscopic redshifts \\
$^e$\,Face-on best-fit effective radii (arcsec)  &  $^j$\,Two-component models (point + exponential) 
\end{tabular}
\end{small}
\end{center}
\leavevmode 
\vspace{0.truecm}
\end{table*}
\end{small}

\section{Observations}
We obtained 102 hours of NIR $J_s,H,$ and $K_s$ imaging in 
the HDF-S ($2\farcm5 \times 2\farcm5$) under excellent
seeing (FWHM$\approx$0$\farcs$46), using ISAAC \citep{Mo97} on the VLT. 
The observations were taken as part of the Faint InfraRed 
Extragalactic Survey \citep[FIRES;][]{Fr00}. 
We combined our data with existing deep optical HST/WFPC2 imaging 
\citep[version 2;][]{Ca00}, in the $U_{300}, B_{450}, V_{606}$ and $I_{814}$ 
bands, 
and we assembled a $K_s$-selected catalog of sources with SExtractor \citep{BA96}. 
Photometric redshifts and rest-frame luminosities were derived by fitting 
a linear combination of empirical galaxy spectra and stellar population models 
to the observed flux points \citep{Ru01,Ru03a}. The reduced images, photometric 
catalog, and redshifts are presented in \citet{La03} and are all available on-line 
at the FIRES homepage\footnote{http://www.strw.leidenuniv.nl/\~{}fires}. 
Furthermore, we obtained optical spectroscopy with FORS1 on the VLT for
some of the sources \citep{Ru03b}. Additional redshifts were 
obtained from \citet{Va02}. As discussed in \citet{Ru01,Ru03a}, our photometric 
redshifts yield good agreement with the spectroscopic redshifts, 
with rms $|z_{spec} - z_{phot}|/(1+z_{spec}) \approx 0.05$ for $z_{spec}>1.4$.

\par
Large disk galaxies in the HDF-South were identified by fitting
exponential profiles convolved with the Point Spread Function (PSF)
to the $K_s$-band images. Six objects at $z>1.4$ have effective 
radii $r_e>3.6$ \hrefkpc, three of which have spectroscopic redshifts.
The mean redshift of the sample is 2.4. We will focus on these large 
galaxies in the remainder of the Letter. The structural properties of 
the full $K_s$-selected sample will be discussed in \citet{T03}.

\section{Rest-frame Optical versus UV Morphology}

The large galaxies are shown in Figure~1.
They  have a  regular morphology in the ISAAC $K_s$-band
($2.2\mu m$),  which probes rest-frame 
optical wavelengths between 5400 and 9000\,\AA. 
In contrast, the WFPC2 $V_{606}$ and $I_{814}$-band morphologies, 
which map the unobscured star-forming regions at rest-frame UV 
wavelengths between 1500 and 3300\,\AA, are irregular with several knots up to 
$\sim 2\arcsec$ ($\sim15$ \hrefkpc) apart, symmetrically distributed 
around the $K_s$-band centers. In a few cases the observed optical light is 
 spatially almost distinct from the NIR. 
\par
As a result of the structure in the WFPC2 imaging, 4 of the large objects
have been split up into two sources by Casertano et al (2003). Fig. 1
shows the corresponding ``segmentation'' map by ``SExtractor'' 
which illustrates how the pixels in each image are allocated to different sources.
The galaxies were not split up when the $K_s$ band image was used to
detect objects. However, the broader PSF in the $K_s$-band image can 
play a role: if we smooth the $I_{814}$-band data to the same resolution, 
we find that SExtractor only splits up 1 galaxy.
\par
Hence, the question remains whether these 4 objects split up in $I_{814}$
are superpositions or whether they are part of larger systems. We tested 
this by subtracting the PSF-matched $I_{814}$-band images from 
the $K_s$-band images. The $I_{814}$-band images were scaled
to the $K_s$-band images to minimize the residuals. 
The residuals are shown in Fig. 1.
In all cases, we find strong positive residuals close to
the centers of the objects as defined in the $K_s$-band, whereas residuals at 
any of the $I_{814}$-band peaks might be expected in case of a chance 
superposition. Furthermore, we performed photometric redshift analyses for subsections of 
the images and found no evidence for components at different redshifts.

\section{Profile fits and sizes}
Next, we fitted simple models convolved with the PSF 
(FWHM$\approx$0$\farcs$46)  
to the two-dimensional surface brightness distributions in the $K_s$-band. 
The images are well-described by a simple 
exponential law over $2-3$ effective radii (galaxy 302, 267, 257 and 657) 
or by a point source plus exponential (galaxy 611 and 494),
where the point source presumably represents the light 
emitted by a compact bulge contributing about 40\% of the light. 
We also derived intensity profiles by 
ellipse fitting. As can be seen in Figure~\ref{fig.a}, most galaxies 
are well described by an exponential.  The central surface brightnesses and
effective radii, enclosing half of the flux of the model profile, 
are corrected to face-on and shown in Table \ref{tab.a}.
The central surface brightness is multiplied with $\sqrt{1-\epsilon}$, 
as an intermediate case between optically thin and optically thick, 
and corrected for cosmological dimming.
\par
The  effective radii (semi-major axes) are surprisingly large, 
$r_e=0\farcs65-0\farcs9$ ($5.0-7.5$ \hrefkpc\ 
in a \lcdm cosmology), comparable to the Milky Way and much larger
than typical sizes of ``normal'' Ly-break galaxies \citep{GSM96,Lo97}.
As might be expected from the previous section, the I-band images
have even larger effective radii. All galaxies have a ``red'' nucleus, 
and the colors become bluer in the outer parts.
\par
Overall, the optical-to-infrared morphologies and sizes are 
strikingly similar to $L^*$ disk galaxies in the local universe,
with red bulges, more diffuse bluer exponential disks and scattered, 
UV-bright star forming regions. Some even show evidence of well-developed 
grand-design spiral structure. 
However, the mean central surface brightness of the disks is $1-2$ 
mag higher than that of nearby disk galaxies and the mean rest-frame color
$(U-V)\approx0$  is $\sim1$ mag bluer (c.f. \citealt{Li98}). 
Passive evolution can lead to disks with normal surface brightnesses
at low redshift. Alternatively, the disks are  disrupted later 
by interactions or evolve into S0's, which have higher 
surface brightnesses (e.g., \citealt{B79}).

\section{Spectral Energy Distribution}
The overall SEDs of the galaxies show a large variety.
Four of the galaxies (257, 267, 494 and 657) satisfy conventional U-dropout 
criteria \citep{M96,GD01}. One galaxy is at too low redshift (302) to
be classified as a U-dropout, and one other galaxy is too faint in the
rest-frame UV (611). It has $J_s-K_s > 2.3$, and is part of the population 
of evolved galaxies identified by Franx et al (2003) and van Dokkum et al (2003).
\par
The red colors of the  central components can be due to either dust,
higher age, emission line contamination, or a combination of effects. 
The SEDs show stronger Balmer/4000\,\AA\ breaks in the inner parts than 
in the outer parts (see Fig. 2 for an example).
We derived colors inside and outside of a 0.7 arcsec radius
centered on the $K_s$-band center. The mean differences are 
$\sim0.5$ mag in observed $I_{814}-Ks$ and $\sim 0.2$ mag in rest-frame $U-V$.
Higher resolution NICMOS data are required to address the population differences in more detail.

\figb

\section{Discussion}
We have found 6 large galaxies with characteristics similar to those
of nearby disk galaxies: exponential profiles with large scale lengths,
more regular and centrally concentrated morphologies in the
restframe optical than in the rest-frame UV, and, as a result, red nuclei.
It is very tempting to classify these galaxies as disk galaxies, given
the similarities with low redshift disk galaxies. However, kinematic 
studies are necessary to confirm that the material is in a rotating disk. 
Photometric studies of larger samples are needed to constrain the thickness 
of the disks. We note that simulations of \citet{SN02} can show extended 
structures during a merging or accretion event. The expect duration of this 
phase is short, however, while the disk galaxies comprise a high fraction 
of the bright objects. It is therefore unlikely that a significant fraction 
of the galaxies presented here are undergoing such an event.
\par
The density of these large disk galaxies is fairly low:
over a survey area of $4.7$ arcmin$^2$ and to a magnitude limit of $K_{s,tot} = 22$
they make up 6 out of 52 galaxies at $1.4 \lesssim z \lesssim 3.0$. 
However, they do constitute 6 out of the 12 most rest-frame luminous
galaxies $L_V \gtrsim 6\times 10^{10} h_{70}^{-2} L_{\sun}$ in the 
same redshift range. The comoving volume density is 
$\sim3\times10^{-4}$ \hrefmpc\ at a mean redshift $z\approx2.3$. 
Obviously, larger area surveys are needed to establish the true density.
We note that three of the galaxies only have photometric redshifts, 
one of which (a U-drop galaxy at $z_phot=1.82$) is in the poorly tested 
range $1.4 < z < 2.0$. The volume density of disk galaxies with $r_e>3.6$ 
\hrefkpc\ in the local universe is much higher at $\sim3\times10^{-3}$ 
\hrefmpc\ \citep{dJ96}, although many nearby disks would not be present 
in our high redshift sample because their surface brightness would be too low.
\par
We note that similar galaxies are absent in the very deep Near-IR imaging data
on the HDF-N ( \citealt{Wi96,Di00})
Although notable differences 
between optical and NIR morphologies were reported for two of the 
largest LBGs in the HDF-N, no large galaxies were reported to have
red nuclei and exponential profiles as in the HDF-S. The
two fields are different in other aspects as well. We found earlier 
that the HDF-N is deficient in red sources (e.g.,
\citealt{La03} versus \citealt{Pa01}) and the disk galaxies are 
more luminous in the $H$-band than most of the high-redshift galaxies
found in the HDF-N \citep{Pa01}. The larger number of red galaxies
in the HDF-S (e.g., \citealt{La03}, \citealt{Fr03}) and their 
strong clustering \citep{Da03} 
may indicate that the red galaxies and large disks are both part of the 
same structures with high overdensities, and evolve into the 
highest overdensities at low redshift, i.e. clusters. If this 
is the case, the large disk galaxies may be the progenitor of 
large S0 galaxies in the nearby clusters, which have very similar 
colors as elliptical galaxies (e.g., \citealt{Bo92}).
\par
Finally, we can compare the observed disk sizes to theoretical predictions.
It is often assumed \citep{FE80,Mo98} that 
the disk scale  length is determined by the spin parameter $\lambda$ 
and the circular 
velocity of the virialized dark matter halo \citep{FE80,Mo98}. 
For a \lcdm cosmology, \citet{MMW99} predict that the $z\sim3$  space density 
of large ($r_e \gtrsim 3.6$ \hrefkpc) bright U-dropouts is 
1.1$\times 10^{-4}$ \hrefmpc, whereas we find $\sim2\times 10^{-4}$ \hrefmpc\ 
for our 4 U-drops at a mean $<z>\sim2.4$. 
This difference of a factor of two is  not very significant, given
our low number statistics and small survey volume. The combination of 
sizes and rotation velocities will give much stronger
constraints on these models; it may be possible to measure the
kinematics of some of these large galaxies using NIR spectrographs on 
large telescopes.

\begin{acknowledgements}
We thank the staff at ESO for their hard work in taking these data and
making them available. This research was supported by grants from the
Netherlands Foundation for Research (NWO), the Leids Kerkhoven-Bosscha
Fonds, and the Lorentz Center. GR thanks Frank van den Bosch and 
Jarle Brinchmann for useful discussions.
\end{acknowledgements}

\end{document}